\pdfoutput=1
\documentclass[epjST,a4paper,final,fleqn]{svjour}
\usepackage{graphicx}
\usepackage[utf8]{inputenc}
\usepackage{engord}
\usepackage{xcolor}
\usepackage{amsmath,amssymb}
\usepackage[T1]{fontenc}
\usepackage{siunitx} 
\usepackage{cite} 
\usepackage{hyperref} 
\hypersetup{
	colorlinks=false,
	linkbordercolor=red,
	linkcolor=green,
	pdfborderstyle={/S/U/W 1}
}

\begin{document}

\title{Double transfer through Dirac points in a tunable honeycomb optical lattice}
\subtitle{}

\author{Thomas~Uehlinger\inst{1}\fnmsep\thanks{\email{uehlinger@phys.ethz.ch}} \and Daniel~Greif\inst{1} \and Gregor~Jotzu\inst{1} \and Leticia~Tarruell\inst{1,2} \and Tilman~Esslinger\inst{1} \and Lei~Wang\inst{3} \and Matthias~Troyer\inst{3} }

\institute{Institute for Quantum Electronics, ETH Zurich, 8093 Zurich, Switzerland \and LP2N, Université Bordeaux 1, IOGS, CNRS, 351 cours de la Libération, 33405 Talence, France \and Theoretische Physik, ETH Zurich, 8093 Zurich, Switzerland}

\abstract{
We report on Bloch-Zener oscillations of an ultracold Fermi gas in a tunable honeycomb lattice.
The quasi-momentum distribution of the atoms is measured after sequentially passing through two Dirac points.
We observe a double-peak feature in the transferred fraction to the second band, both as a function of the band gap at the Dirac points and the quasi-momentum of the trajectory. 
Our results are in good agreement with a simple analytical model based on two successive Landau-Zener transitions. 
Owing to the variation of the potential gradient over the cloud size, coherent Stückelberg oscillations are not visible in our measurements. This effect of the harmonic confinement is confirmed by a numerical simulation of the dynamics of a trapped 2D system.
} 

\maketitle

\section{Introduction}

The transport properties of a quantum particle in a periodic potential
depend crucially on the band structure of the system. A prime example is graphene, where
remarkable transport behaviour results from the presence of two Dirac
points in the band structure of its honeycomb lattice
\cite{CastroNetoRMP09}. At a Dirac point, two energy bands intersect
linearly and a quantum particle behaves as a relativistic Dirac fermion.
Dirac points appear naturally in a number of condensed matter systems
besides graphene, including nodal points in \emph{d}-wave superconductors
or the surface states of three-dimensional topological insulators
\cite{HasanRMP10}. The transport properties associated to them can also be
studied in artificially engineered systems using, for example, electrons in
molecular graphene \cite{GomesNature12}, microwave fields in meta-material
structures \cite{KuhlPRB10} or ultracold atoms in optical lattices.

Quantum gases in optical lattices have recently emerged as an attractive
system for exploring Dirac points in a highly tunable environment. By
loading a Bose-Einstein condensate into the excited bands of a one
dimensional bichromatic optical lattice, the one-dimensional analogue of a
Dirac point was realized and relativistic effects such as Zitterbewegung
\cite{KlingPRL10} and Klein tunnelling \cite{SalgerPRL11} were observed.
Using a degenerate Fermi gas trapped in a two-dimensional
honeycomb optical lattice, tunable Dirac cones were realized and the
merging of Dirac points could be explored for the first time
\cite{TarruellNature12}.

Another key feature of ultracold atoms is the possibility to perform
transport studies in a dissipation-free environment, and to hence
explore regimes which are difficult to access in solid state samples. For
example, in the presence of a constant force, Bloch oscillations --the
oscillatory motion of a particle in a periodic structure-- can be observed
\cite{DahanPRL96}. In the presence of Dirac points, Bloch-Zener
oscillations, a characteristic sequence of Bloch oscillations and
Landau-Zener transitions between the two crossing energy bands, appear
instead \cite{BreidNJP06}. Their study allows for a precise characterization of the
underlying band structure, giving access to properties such as the energy
splitting between the bands \cite{KlingPRL10} and the position of the
Dirac points \cite{TarruellNature12}.

In this paper, we present a systematic study of Bloch-Zener oscillations
of an ultracold Fermi gas loaded into a tunable honeycomb lattice,
focusing on quasi-momentum trajectories for which the two Dirac points
are successively passed during the Bloch cycle. In the presence of a small
energy gap at these band crossings each of them acts as an atomic beam
splitter, where partial Landau-Zener tunnelling between the two energy bands
occurs. Therefore, in this configuration the atomic wave-packet is
successively splitted and then recombined during the Bloch cycle. Within the atomic cloud and when scanning system parameters we observe different behaviors
depending on whether the probability of transfer at each band crossing close to a Dirac point is below
or above $1/2$.

After
presenting our experimental setup in Section \ref{section:setup}, we explore the
effect of the double Landau-Zener transition in Section \ref{section:doublepass}.
The experimental sequence is described in detail in Section \ref{subsection:sequence}.
The fraction of atoms transferred to the second energy band is
recorded as a function of the transverse quasi-momentum and the lattice
geometry in Sections \ref{subsection:LZ_transfers} and \ref{s:slope}, and as a function of the
sublattice offset in Section \ref{subsection:gap}. We find good agreement with a simple analytical model based on two
successive Landau-Zener tunnelling events. 
Despite the fact that the splitting process at the
Dirac points is expected to be coherent, no interference fringes are
observed, an aspect which is studied in detail in Section \ref{subsection:Stueckelberg}. We
conclude in Section \ref{section:conclusion}.

\section{Experimental setup}\label{section:setup}

\begin{figure}[ht]
	\centering
  \includegraphics[width=10cm]{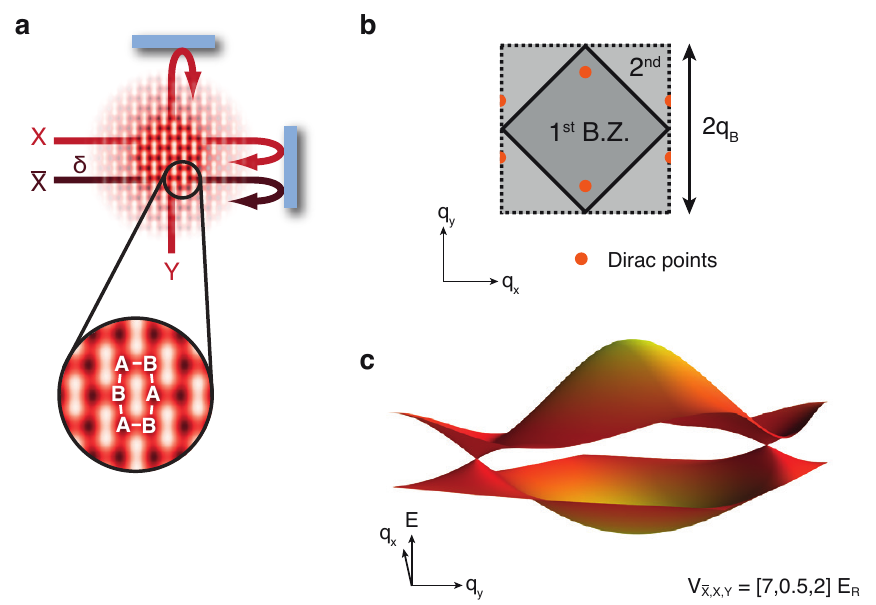}

  \caption{{\bf Experimental scheme.} {\bf a} Three retro-reflected laser beams give rise to an optical potential of honeycomb geometry. The lattice has a unit cell consisting of two sites $A$ and $B$. {\bf b} The corresponding \engordnumber{1} and \engordnumber{2} Brillouin zones are shown, with the two Dirac points depicted as orange circles. Here $q_B=2
\pi/\lambda$ denotes the Bloch wave vector. {\bf c} Band structure of a honeycomb lattice for typical parameters.}
   \label{fig:fig1}
\end{figure}

Recently, optical lattices with complex geometries have been implemented in various experimental setups~\cite{StrableyPRA06,FoellingNature07,SalgerPRL07,WirthNaturePhys11,SoltanPanahiNaturePhys11,JoPRL12}.
The tunable optical lattice used in this setup is created in the same
manner as described in Ref. \cite{TarruellNature12}, combining in the $x$-$y$ plane three retro-reflected
laser beams of wavelength $\lambda=1064$ nm (see Fig.~\ref{fig:fig1}a). The two
beams $X$ and $Y$ are arranged at an angle of $90.0(1)^\circ$ with respect to each other
and have the same frequency and polarization. They hence create a chequerboard
lattice with a distance of $\lambda/\sqrt{2}$ between potential minima. An additional
beam $\overline{X}$, propagating in the same direction as $X$ but with a frequency
detuning $\delta \approx 0.4\,\mathrm{GHz}$, gives rise to an independent standing wave of spacing $\lambda/2$.

The overall potential is given by

\begin{eqnarray} V(x,y) & = & -V_{\overline{X}}\cos^2(k
x+\theta/2)-V_{X} \cos^2(k
x)\nonumber\\
&&-V_{Y} \cos^2(k y)-2\alpha \sqrt{V_{X}V_{Y}}\cos(k x)\cos(k
y)\cos\varphi,\label{eqlattice}
\end{eqnarray}
where $V_{\overline{X}}$, $V_{X}$ and $V_{Y}$ denote the lattice depths corresponding
to each single laser beam (see Appendix~\ref{s:harmonictrap}), $k=2\pi/\lambda$ and the measured visibility of the interference
pattern $\alpha$ is $0.90(5)$. The phase $\varphi$ is stabilized to $0.00(3)\pi$, whilst
$\theta$ can be continuously adjusted by tuning $\delta$ and is set to values close to $\pi$.
The beams additionally cause a weak harmonic confinement for the atoms in all spatial directions,
as described in Appendix \ref{s:harmonictrap}.

The geometry of the lattice can be changed by adjusting the intensities of the three laser beams.
In the following we focus on the honeycomb configuration, shown in Fig.~\ref{fig:fig1}a.
The lattice has a unit cell consisting of two sites, labelled $A$ and $B$. It gives
rise to a band structure where the two lowest bands are connected by two Dirac points,
but are well separated from higher energy bands.
Since the angle between the primitive lattice vectors is set to $90^\circ$, the first Brillouin
zone (B.Z.) has a square shape, and the Dirac points are located inside it (see Figs. \ref{fig:fig1}b and c).
In the $q_x$ direction, the reflection symmetry of the lattice fixes the Dirac points to
the $q_x=0$ line. Their position in the $q_y$ direction as well as the linear slope
of the band structure in the vicinity of the Dirac points is determined by the strength
of the horizontal and vertical tunnel couplings.
The tunnelling can be controlled through the intensities of the laser beams forming the lattice.
For example, increasing $V_{\overline{X}}$ will mainly reduce the tunnelling in the $x$ direction.

Additionally, tuning the phase $\theta$ away from $\pi$ creates an energy offset $\Delta_{AB}$ between
sites $A$ and $B$ and hence breaks the inversion symmetry of the lattice. This causes
a coupling between the previously orthogonal levels crossing at the Dirac points,
and thus opens up a gap $\Delta$ there, which is proportional to~$\Delta_{AB}$.

\section{Double transfer through Dirac points}\label{section:doublepass}

\subsection{Experimental sequence}\label{subsection:sequence}

We explore the band structure of our system using Bloch-Zener oscillations.
The starting point of each measurement, as described in Ref. \cite{TarruellNature12},
is an ultracold cloud of $N\simeq50,000$ fermionic $^{40}$K~atoms in the $m_F=-9/2$ Zeeman level of the $F_\mathrm{hf}=9/2$ hyperfine manifold.
The gas is loaded into the lowest band of the tunable lattice. We set $V_X/E_R=0.28(1)$ and $V_Y/E_R=1.8(1)$ unless otherwise stated. 
Here $E_R=h^2 /2m\lambda^2$, $m$ denotes the mass of $^{40}$K and $h$ is the Planck constant.
Along the third spatial direction~$z$ the atoms are only confined by a weak harmonic trap.

\begin{figure}[ht]
	\centering
  \includegraphics[width=9cm]{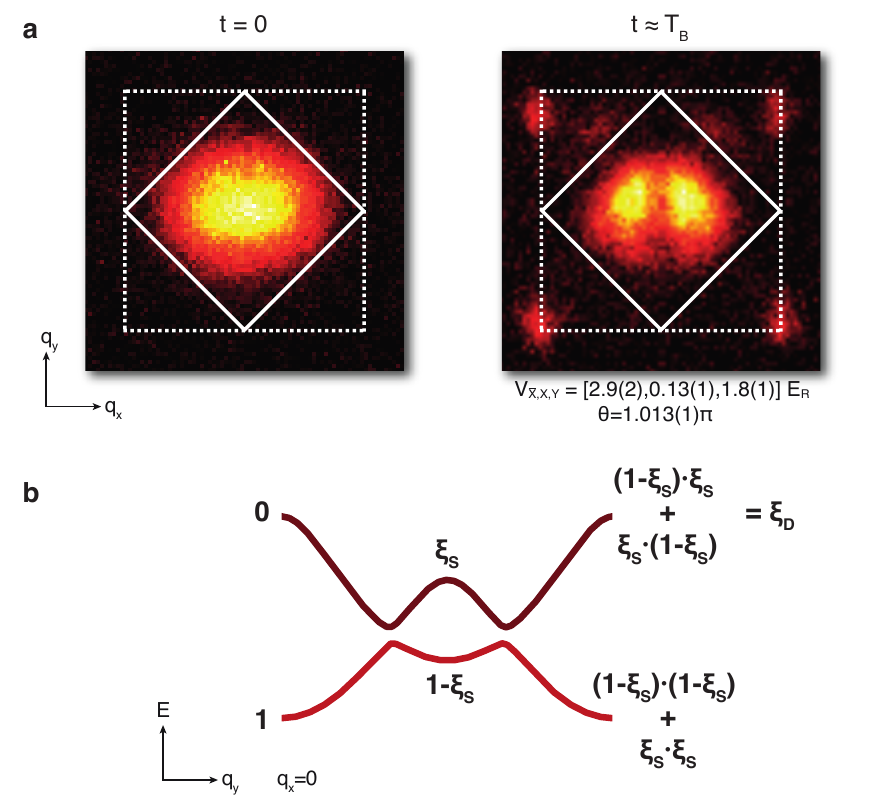}
  \caption{{\bf Probing the Dirac points.} {\bf a} Absorption images of the quasi-momentum distribution of the atomic cloud after preparation ($t=0$) and after performing one Bloch oscillation cycle ($t \approx T_\mathrm{B}$). Colour scale: white indicates a high density of atoms, black indicates no atoms. {\bf b} Cut through the band structure at $q_x=0$. 
  During one Bloch cycle, the atoms pass through two band crossings. For $q_x=0$ the band crossings coincide with the two Dirac points. Neglecting the phase evolution (see Section \ref{subsection:LZ_transfers}), each band crossing has a transfer probability $\xi_S$. }
  \label{fig:fig2}
\end{figure}

We apply a force with characteristic energy $E_{B}/h= F\lambda /2h=95(1)\,\mathrm{Hz}$ to the atoms along the $y$~direction using a magnetic field gradient, leading to Bloch oscillations \cite{DahanPRL96} with a period of $T_B=h/E_B$.
After \SI{11.5}{\milli \second} (corresponding to roughly one Bloch oscillation cycle) the force is turned off.
The optical lattice is then linearly ramped down in \SI{500}{\micro \second} and the gas is allowed to expand freely for \SI{15}{\milli \second} time of flight before an absorption image is taken. This band-mapping technique \cite{KoehlPRL05,KastbergPRL95,Esslinger10} is used to measure the $z$-integrated quasi-momentum distribution of the atoms in the different energy bands.
The final absorption image hence gives information about where in the Brillouin zone interband transfers have occurred.
This can be seen in Fig. \ref{fig:fig2}a, where for $t \approx T_B$ atoms appear in the \engordnumber{2} B.Z. and are missing at the corresponding quasi-momenta in the \engordnumber{1} B.Z.  

\subsection{Landau-Zener transitions}\label{subsection:LZ_transfers}

During one Bloch cycle the cloud probes the dispersion relation of the two lowest bands along the vertical $q_y$ direction, which contains two avoided linear band crossings, see Fig. \ref{fig:fig2}b. For $q_x=0$, the band crossings coincide with the two Dirac points.
Neglecting the harmonic trap, one can treat trajectories for different $q_x$ as independent. 
We attribute a transfer probability $\xi_S$ to a single band crossing along a $q_x$-trajectory, see Fig. \ref{fig:fig2}b. Neglecting the phase evolution and resulting interference of the two bands between the crossings (see Section \ref{subsection:Stueckelberg}), the probability for ending up in the higher band after passing the second crossing can be obtained by simply multiplying the probabilities for passage at the two crossings and summing up the two possible paths:
\begin{equation}
\xi_D=\left(1-\xi_S\right)\xi_S + \xi_S\left(1-\xi_S\right) = 2\xi_S\left(1-\xi_S\right).
\end{equation}
The total transfer $\xi_D$ therefore has a maximum value of $1/2$ for a Landau-Zener transition probability of $\xi_S=1/2$.

\begin{figure}[ht]
	\centering
  \includegraphics[width=1.0\textwidth]{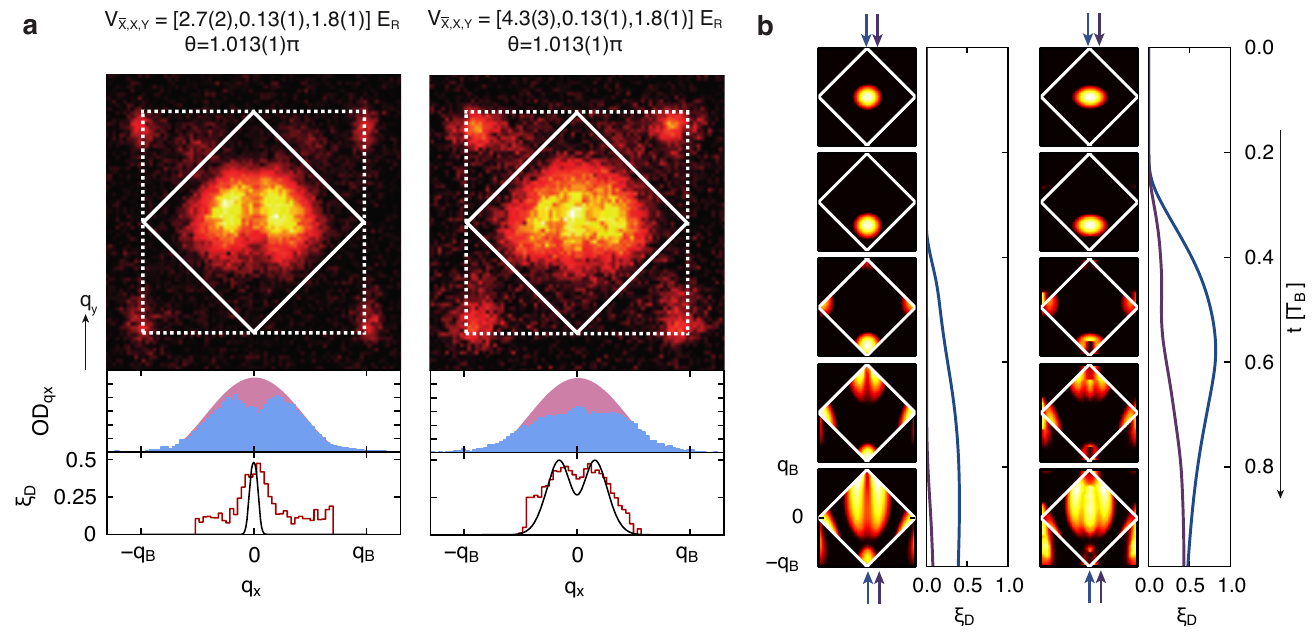}
 
  \caption{{\bf Transfer for different quasi-momentum-trajectories.} {\bf a} Transfer for two sets of lattice parameters. Left: Maximum transfer is observed for a central trajectory. Right: Maximum transfer happens to the left and right of the central trajectory. The plots depict the line sums along $q_y$ of the measured optical densities as a function of $q_x$ (blue area) and the expected line sums without transfer (purple area). From this the transfer $\xi_D(q_x)$ is calculated and compared to the prediction of the analytical model (red and black curve). {\bf b} Numerical simulation of a trapped 2D system with $N=256$ atoms: quasi-momentum distribution during one Bloch cycle for two exemplary situations. The population of the \engordnumber{2}~B.Z. is plotted for two different $q_x$-trajectories (position indicated by the coloured arrows). For the simulation parameters see Appendix \ref{s:simulation}.
  }
  \label{fig:fig3}
\end{figure}

 Depending on the lattice parameters we either observe a single or a double slit of missing atoms in the lowest Brillouin zone after one full Bloch cycle, see Fig.~\ref{fig:fig3}a. The appearance of a double-peak feature in the transfer fraction $\xi_D(q_x)$ is a direct consequence of the double transition through the two band crossings: for the lattice parameters on the left panel of Fig. \ref{fig:fig3}a, $\xi_S<1/2$ at the central $q_x=0$ line, whereas $\xi_S>1/2$ for the right panel due to the different band structure, as will be detailed below. 
 The single transfer probability $\xi_S$ generally decreases for quasi-momenta away from the center due to the increased gap at the crossing. Therefore, the maximum total transfer (i.e. $\xi_D=1/2 \Leftrightarrow \xi_S=1/2$) is reached at a finite value of $q_x$ for the latter case. This gives rise to the double-peak structure. 
 It appears due to the transfer probability $\xi_S$ being close to 1 at the two linear band crossings at the Dirac points.

For a quantitative treatment, we perform a line sum of the optical density (OD) in the \engordnumber{1} B.Z. (blue area in the figure). The initial cloud profile obtained from a fit to the quasi-momentum distribution after a Bloch cycle for a set of lattice parameters not giving rise to Dirac points is shown in purple.
From these two profiles, the experimental transfer fraction $\xi_D(q_x)$ (red curve below) can be obtained. 
As can be seen in the figure, the transfer is peaked at $q_x=0$ for the left situation, whilst we observe a double-peak (with a dip at $q_x=0$) for the situation on the right.

The exact value of $\xi_S$ depends on the gap at the band crossing and the slope of the dispersion relation in the $y$~direction far away from the crossing.
To evaluate the transfer probability $\xi_S$ we use a simplified effective Hamiltonian well suited for describing the dispersion relation close to the two Dirac points, an approach previously worked out in detail in Refs. \cite{MontambauxEPJ09,LimPRL12},
\begin{equation}
H = \left( \begin{array}{cc}
0 & \Delta_*+\frac{\hbar^2 q_y^2}{2m^*}-i c_x q_x \\
\Delta_*+\frac{\hbar^2 q_y^2}{2m^*}+i c_x q_x & 0
\end{array} \right).
\end{equation}
Here $\Delta_*=-\frac{c_y^2 m^*}{2\hbar^2}$ denotes the merging gap, $m^*$ the effective mass along $q_y$ at the saddle point between the two Dirac points and $c_x$ and $c_y$ the slopes of the dispersion relation along the different directions at the Dirac point, with $\hbar=h/2\pi$ being the reduced Planck constant. Their values are directly related to the lattice parameters~\cite{LimPRL12}. Denoting the gap at the Dirac points with $\Delta$, the transfer probability is then
\begin{equation}
\xi_S=\mbox{exp}\left(-\pi\frac{c_x^2q_x^2+\Delta^2/4}{c_y F}\right). 
\end{equation}
The theoretical prediction of this simple analytical model is shown in the black curve in the bottom panel of Fig. \ref{fig:fig3}a. The main features of the experimental momentum-resolved transfer fractions obtained from the atomic distribution in the lowest band are captured by theory. Deviations are possibly due to the finite resolution of the band-mapping technique as well as the uncertainties in the calibration of $V_{\overline{X}}$.

The simple picture of two independent Landau-Zener transitions is confirmed when performing a numerical time-evolution of a trapped 2D lattice system in a tight binding limit, see Fig. \ref{fig:fig3}b (details of the simulation see Appendix \ref{s:simulation}). The transfer fraction at $q_x=0$ increases with time on the left panel as the atoms pass the two Dirac points. On the right, however, after having passed through the first crossing, the transferred fraction decreases again. In contrast, the transferred fraction increases monotonously in both cases for quasi-momenta away from the center. 

\begin{figure}[ht]
	\centering
  \includegraphics[width=10cm]{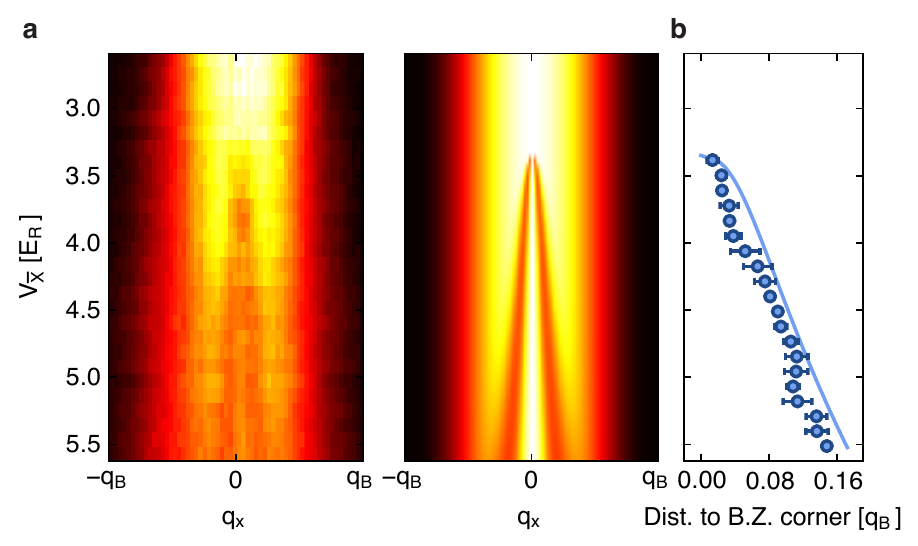}
 
  \caption{{\bf Transfer for different lattice parameters.} {\bf a} Experimental (left) and theoretical (right) quasi-momentum distribution $\xi_D(q_x)$ in the \engordnumber{1} B.Z. after one full Bloch cycle integrated along the $q_y$ direction for different values of $V_{\overline{X}}$.  The experimental data is the average of three consecutive measurements.
  The theoretical quasi-momentum distributions are calculated under the assumption of a fixed $q_x$-distribution for the initial atomic cloud, which is obtained by a fit to the density profile at $V_{\overline{X}}/E_R=2.5$ in the left panel. At this point the band structure does not contain Dirac points. {\bf b} Extracted position of maximum transfer along $q_x$ using the distribution in the \engordnumber{2} band. Values and
error bars denote the mean and standard deviation of three consecutive
measurements The solid line shows the theoretical expectation of the maximum transfer position using the simple analytical model without free parameters. 
  The experimental data was taken with $\theta$ set to $1.000(1) \pi$.
 }

  \label{fig:fig4}
\end{figure}

\subsection{Changing the slope of the Dirac cones}\label{s:slope}

As clearly visible in Fig. \ref{fig:fig3}, the transfer depends on the precise lattice parameters. Since these values are freely tunable in our experiment, the dependency on the lattice structure can be systematically investigated. Fig. \ref{fig:fig4}a shows such a scan versus $V_{\overline{X}}$ along with the theoretical expectation using the simple analytical model. 
For $V_{\overline{X}}/E_R<3.4$ there are no Dirac points in the band structure and thus no significant transfer is observed (no missing atoms). For $V_{\overline{X}}/E_R>3.4$ two Dirac points are present in the band structure, leading to the formation of a progressively more pronounced double-peak feature as $V_{\overline{X}}$ gets larger. This is caused by the overall increased single transfer probability $\xi_S$ for larger $V_{\overline{X}}$ due to the deformation of the Dirac cones, shifting the points of maximum transfer $\xi_D$ further apart.
The position along $q_x$ in quasi-momentum space for maximum transfer can in fact be obtained by extracting the peak position of the atomic distribution in the \engordnumber{2} B.Z. after taking a line sum along $q_y$. The results of this procedure are shown in Fig. \ref{fig:fig4}b, which are in good agreement with the calculated position obtained from the simple analytical model. 

\subsection{Opening a gap at the Dirac points}\label{subsection:gap}

The presence of a transition regime from a single- to a double-peak feature in the momentum resolved transfer $\xi_D(q_x)$ depends on the gap at the Dirac points, which is controlled by the energy offset between neighbouring sites. For a vanishing gap only double-peak features appear, as the single transfer in the center is always close to $1$. This is the case for the data presented in Fig. \ref{fig:fig4}, as $\theta \approx \pi$ for this dataset. In contrast, an overly large gap merely leads to single-peak profiles with a very low overall transfer fraction, as the transfer in the center is already far below $1/2$. 
Only if the gap is set to an intermediate value, a transition from a single to a double-peak profile occurs for increasing $V_{\overline{X}}$, which is the case for the parameters of the measurements in Fig.~\ref{fig:fig3}.

\begin{figure}[ht]
	\centering
  \includegraphics[width=9cm]{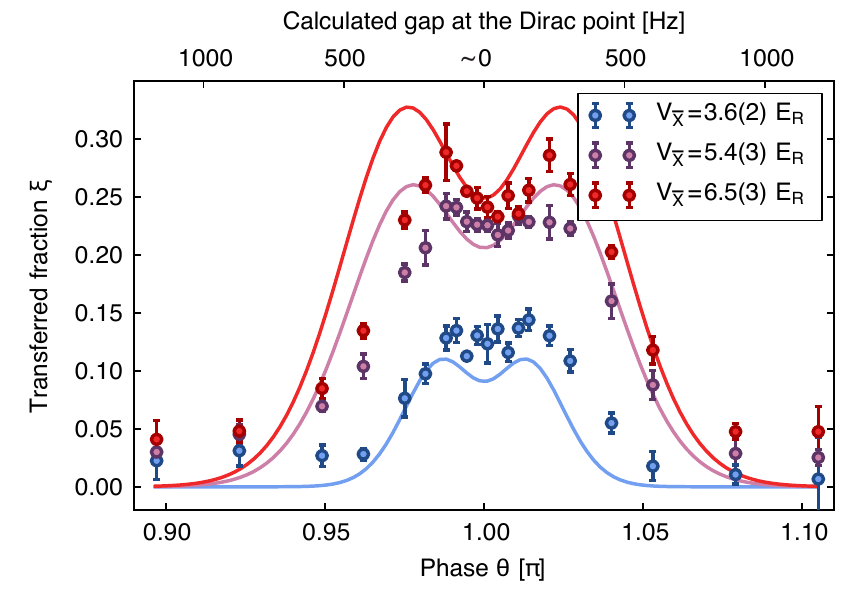}
 
  \caption{{\bf Scanning the gap at the Dirac point.} The total transfer fraction $\xi$ is measured for three different lattice depths $V_{\overline{X}}$ versus the phase $\theta$, which directly controls the energy offset between neighbouring sites. The gap at the Dirac point is obtained from a band structure calculation \cite{TarruellNature12}. Values and
error bars denote the mean and standard deviation of five consecutive
measurements. The solid lines show the theoretical predictions from the simple analytical model without fitting parameters including the integration over $q_x$.}
  \label{fig:fig5}
\end{figure}

From this simple picture we expect the total 3D averaged transfer fraction $\xi$ to exhibit a dip as a function of the gap $\Delta$ at the Dirac point, since a large fraction of the atoms is located around $q_x=0$. A scan of the transfer fraction versus $\Delta$ for three different lattice configurations is shown in Fig. \ref{fig:fig5}, where the transfer $\xi$ for the entire atomic cloud was obtained by comparing the number of atoms after one full Bloch cycle in the two lowest bands \cite{TarruellNature12}. We observe a clear double-peak structure, which reduces for smaller values of $V_{\overline{X}}$. This is caused by the decreased slope $c_y$ and increased slope $c_x$ of the dispersion relation close to where the two Dirac points merge and annihilate, leading to an overall reduction of $\xi_S$ and thus to a less pronounced double-peak feature.
The main features of the experimental results agree with the predictions of the simple analytical model using an integration over the entire cloud (see Appendix \ref{sec:integration}). 
To calibrate the laser detuning $\delta$ for which $\theta=\pi$ we used Bloch oscillations along the $x$ direction \cite{TarruellNature12}. The slight shift of the symmetry axis for the data presented here is smaller than an estimate for a possible systematic error\footnote{The estimate is based on an comparison to an independent calibration of the $\theta=\pi$ point using Raman-Nath diffraction on a $^{87}$Rb Bose-Einstein condensate \cite{TarruellNature12}.}.

The comparison to theory allows to give an upper estimate for a possible small residual gap at the Dirac points. For the dataset where $V_{\overline{X}}=3.6\,E_R$ a gap of $140\,\mathrm{Hz}$ would already lead to a vanishing of the double-peak structure. From this we conclude that, if there is any residual small gap at the Dirac points, for example caused by the finite size of the system, it is significantly smaller than $140\,\mathrm{Hz}$. This is about a factor of 30 smaller than the bandwidth.

\subsection{Stückelberg interference}\label{subsection:Stueckelberg}

\begin{figure}[ht]
	\centering
  \includegraphics[width=9cm]{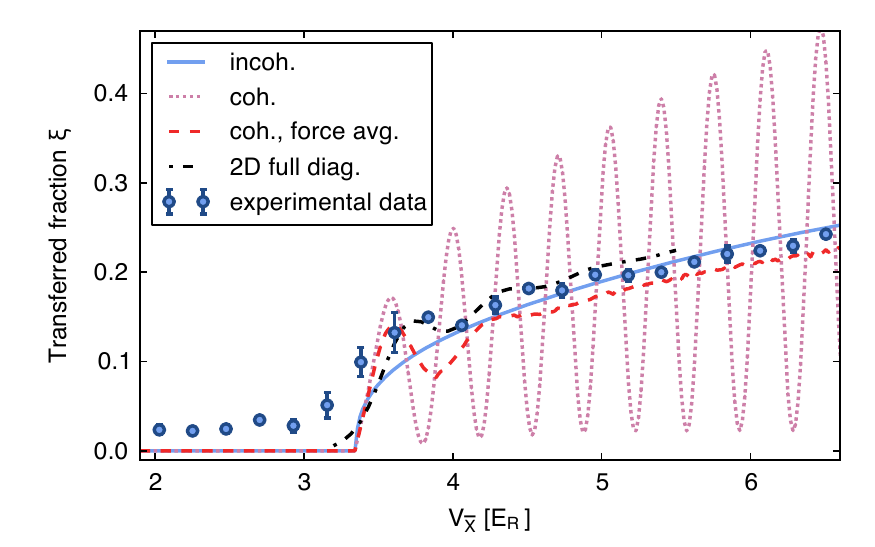}
  \caption{{\bf Contrast of Stückelberg interference.} Total transfer fraction $\xi$ versus $V_{\overline{X}}$ for $\theta\ = 1.000(1) \pi$. Values and
error bars denote the mean and standard deviation of three consecutive
measurements. The lines show the theoretical prediction for the fully integrated transfer fraction using the simple incoherent analytical model (blue solid line) and the extended model including the Stückelberg correction factor (purple dotted line). The red dashed line includes a gaussian distribution of forces with a width of $60\,$Hz resulting from the harmonic trapping potential. The result from the numerical time-evolution of the trapped 2D system for $N=256$ atoms is shown as the black dash-dotted line.}
  \label{fig:fig_stueckelberg}
\end{figure}

A full description of the system studied here would necessarily take coherence into account. The two sequential band crossings along one $q_x$-trajectory would therefore effectively realize a Stückelberg interferometer \cite{StueckelbergHPA32,ShevchenkoPR10} with an associated dynamical phase $\phi$.
This phase depends both on the energy difference between the upper and the lower path (see Fig. \ref{fig:fig2}b) and the time spent in between the two crossings, and therefore on the lattice geometry and the applied force.
In the simple analytical model used so far this phase has been neglected.
In fact, it can be taken into account by multiplying the transfer fraction with a correction factor $2\mbox{cos}^2(\phi/2+\phi_d)$, where the dynamical phase~$\phi$ and transfer phase~$\phi_d$ are defined in \cite{LimPRL12}. 
Fig. \ref{fig:fig_stueckelberg} shows a comparison of the total 3D averaged transfer $\xi$ versus $V_{\overline{X}}$ for the simple incoherent model (blue line) and the extended coherent model including this correction factor (purple dotted line). The experimental data agrees very well with the incoherent model, whereas the oscillatory behaviour as predicted by the extended model cannot be observed.

For a correct treatment of the Stückelberg interference, the variation of the effective force~$F$ due to the harmonic confinement over the cloud size has to be taken into account. Using an estimate for the cloud size based on the trapping frequencies of approximately $\SI{30}{\micro \meter}$ one calculates a variation of the effective force over the cloud on the order of $60\,\mathrm{Hz}$. We include this effect by assuming a gaussian distribution of forces over the entire sample, leading to different Stückelberg oscillation frequencies and thus to a reduced contrast of the oscillations. The result is shown in Fig. \ref{fig:fig_stueckelberg}, where the force averaged curve (red dashed line) is in close agreement with the experimental data. This picture is confirmed by the results of the numerical simulation of the trapped lattice system (black dash-dotted line), which already shows a comparable reduction in contrast of the oscillations for a strictly 2D system.
While additional effects, such as the variation of the lattice depths along the third spatial axis $z$, are smaller than the effect identified above, they probably cause the remaining loss of contrast leading to no visible oscillations in the experimental data. Therefore it can be safely assumed that the simple incoherent version of the analytical model is sufficient for comparison with the experiment. 

\section{Conclusion}\label{section:conclusion}

We have investigated the intricate dynamics of interband transitions arising from the presence of two Dirac points.
Subsequent transfer through two linear band crossings leads to the appearance of a distinct feature: double peaks in the probability of finding atoms in the higher band after one Bloch oscillation, both as a function of sublattice energy offset as well as of quasi-momentum $q_x$. We find good agreement
with a theory without free parameters, based on a universal Hamiltonian describing the vicinity of the Dirac points \cite{MontambauxEPJ09,LimPRL12}. The absence of Stückelberg interference was successfully explained by large differences in the acquired Stückelberg phases due to the inhomogeneity of the force applied to the atoms -- an effect of the harmonic trapping potential. This explanation is confirmed by numerical simulations of a trapped 2D lattice system.

The momentum-resolved measurement of two sequential Landau-Zener transitions has been shown to be a sensitive probe of the band structure of our system. In this paper, the effect of varying the geometry of the lattice and the energy offset between sublattices was characterized in this way. In future work the influence of factors which are more challenging to model theoretically, such as impurities, interactions, finite size or different types of edges could be studied in the same manner. Furthermore,
cold atoms in optical lattices may become a useful tool for investigating open questions in relativistic quantum mechanics using the linear dispersion relation surrounding a Dirac point.

\paragraph{Acknowledgments} 
We acknowledge SNF, NCCR-MaNEP, NCCR-QSIT, and SQMS (ERC advanced grant) for funding. 
LW and MT have been supported by a grant from the Army Research Office through the DARPA OLE program. The numerical simulations of the trapped lattice system were run on the Brutus cluster at ETH Zurich and the Monte Rosa cluster at the Swiss National Supercomputing Centre~(CSCS). 

\appendix

\section{Numerical simulation}\label{s:simulation}

\begin{figure}[hbtp]
\centering
  \includegraphics[width=4cm]{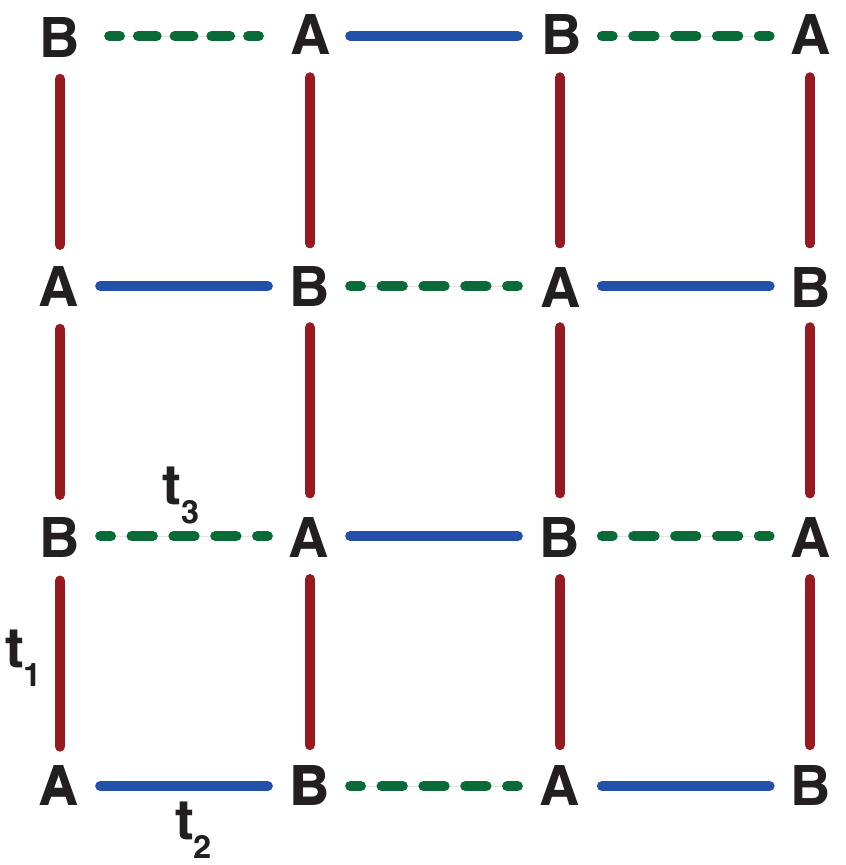}
\caption{{\bf The tight binding model.} Each unit cell contains two sites ($A$ and $B$). The hopping amplitudes for the red, blue and green bonds are $t_{1},t_{2}$ and $t_{3}$ respectively. For $t_3 = 0$ the model reduces to a brick-wall lattice, which is topologically equivalent to the honeycomb lattice. }
\label{fig:lattice}
\end{figure}

We describe the optical lattice with the following tight-binding Hamiltonian \cite{LimPRL12},  

\begin{eqnarray}
H_\mathrm{lattice} & = & -t_{1}\sum_{\langle ij\rangle \in r} c^{\dagger}_{i} c_{j} -t_{2}\sum_{\langle ij\rangle \in b} c^{\dagger}_{i} c_{j} -t_{3}\sum_{\langle ij\rangle \in g} c^{\dagger}_{i} c_{j}  \nonumber\\ & & + \frac{\Delta_{AB}}{2} \sum_{i\in A} n_{i} - \frac{\Delta_{AB}}{2} \sum_{i\in B} n_{i}
\label{eq:TBHam}
\end{eqnarray}
where the hopping amplitudes of the red, blue and green bonds (r,b,g) are $t_{1},t_{2}$ and $t_{3}$ respectively, see Fig. \ref{fig:lattice}. The lattice consists of two sites per unit cell.
$\Delta_{AB}$ is the staggered onsite energy on the $A$ and $B$ sublattices.  The tight-binding parameters $t_{1},\, t_{2},\, t_{3}$ and $\Delta_{AB}$ are extracted from fits of the tight-binding Hamiltonian to the band structure calculated for the potential given in Eq.~\ref{eqlattice}. The real-space Hamiltonian~(Eq.~\ref{eq:TBHam}) is Fourier transformed to momentum space to get a two-by-two matrix $\hat{H}_\mathbf{k}$. It is diagonalized by $\hat{U}^\dagger_\mathbf{k} \hat{H}_\mathbf{k} \hat{U}_\mathbf{k}= E_\mathbf{k}$. When $\Delta_{AB} =0$ and $t_{2}+t_{3}<2t_{1}$, the tight-binding model features Dirac points \cite{HasegawaArXiv12}.

In the experiment there is an additional harmonic trapping potential 

\begin{equation}
H_{\mathrm{trap}}=\sum_{i} (\gamma_{x} x_{i}^{2}+ \gamma_{y} y_{i}^{2})n_{i}\, ,  
\end{equation}
where $x_{i},y_{i}$ are the spatial coordinates of the  $i$-th site. They are measured in units of $\lambda/2$.
 $\gamma_{x(y)}=\frac{1}{2} m\omega_{x(y)}^{2}(\lambda / 2)^{2}$ are the strengths of the harmonic confinement along the $x$ ($y$) direction. They are determined from the intensities of the laser beams (see Appendix \ref{s:harmonictrap}). 

To simulate the Bloch-Zener oscillation experiment, we first solve the ground state~$|\Psi\rangle$ of $N=256$ spinless fermions of $H_\mathrm{lattice}+H_\mathrm{trap}$, then apply a linear gradient field $H_\mathrm{field}=F \lambda / 2 \sum_{i}  y_{i} n_{i}$ to the system and evolve the wave function

\begin{equation}
|\Psi(t)\rangle = e^{-i(H_\mathrm{lattice}+H_\mathrm{trap}+H_\mathrm{field})\frac{t}{\hbar}}|\Psi \rangle \, .
\end{equation}
The time evolution is performed with the Lanczos algorithm \cite{Park} with $200$ Lanczos vectors. The number of lattice sites ($2\times 200^2$) is chosen such that the cloud does not touch the boundary during the Bloch oscillation. The time step is $0.05T_{B}$. 
At each time step, we measure the density matrix 
$
  \rho_{i j}  =  \langle \Psi(t)| c_j^{\dagger} c_i|\Psi(t)\rangle 
$ of the system. 

To extract momentum distributions, we reshape $\rho_{ij}$ into $\rho_{IJ}^{ab}$, where $I,J$ are indices for the unit cell, and $a,b$ are sublattice indices. A Fourier transform with respect to $I,J$ gives $\rho_\mathbf{k}^{ab}=\hat{\rho}_{\mathbf{k}}$. Applying the unitary transformation  $\hat{U}^\dagger_\mathbf{k} \hat{\rho}_\mathbf{k}\hat{U}_\mathbf{k}$, the two diagonal elements become the band filling $n_{\mathbf{k},\mathrm{lower}}$ and $n_{\mathbf{k},\mathrm{upper}}$. 

\begin{figure}[ht]
	\centering
  \includegraphics[width=12cm]{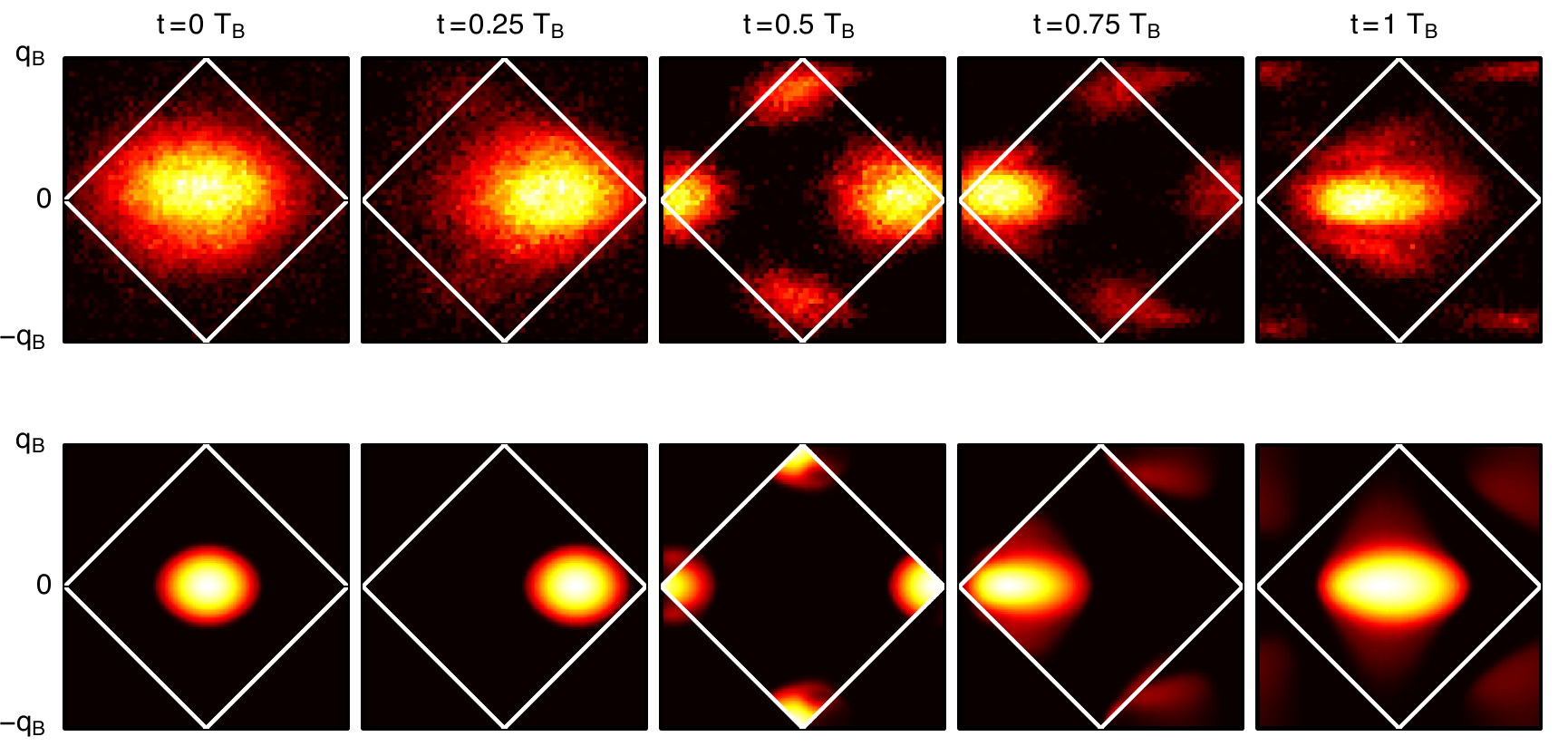}
  \caption{{\bf Time-resolved Bloch oscillations.} Comparison of the quasi-momentum distribution for Bloch oscillations resulting from a force pointing along $x$ in the experiment (top) and in the numerical simulation of a 2D trapped system for $N=256$ atoms (bottom).
  }
  \label{fig:fig_movie}
\end{figure}

As a demonstration of the method, a comparison of one time-resolved Bloch oscillation cycle for a force pointing along $x$ is shown in Fig. \ref{fig:fig_movie}. As can be seen in the figure, the simulation captures the essential features of the experimental data.

\subsection*{Numerical simulation parameters}

\begin{description}

\item[{\bf Fig. \ref{fig:fig3}b.}]
  Left simulation: $t_{1,2,3}/h=[589,969,184]\,\mathrm{Hz}$, $\gamma_{x,y}/h=[0.09,0.28]\,\mathrm{Hz}$, $\Delta_{AB}/h=20\,\mathrm{Hz}$, $F\lambda/2h=63 \, \mathrm{Hz}$, right: $t_{1,2,3}/h=[620,525,131]\,\mathrm{Hz}$, $\gamma_{x,y}/h=[0.09,0.28]\,\mathrm{Hz}$, $\Delta_{AB}/h=17\,\mathrm{Hz}$, $F\lambda/2h=63 \, \mathrm{Hz}$.
  
\item[{\bf Fig. \ref{fig:fig_stueckelberg}, black dash-dotted curve.}]
$t_{1,2,3}$ as obtained from a fit to the calculated band structure for the given lattice parameters. $\gamma_{x,y}$ as deduced from the lattice parameters, see Appendix \ref{s:harmonictrap}. $\Delta_{AB}/h \approx 20\,\mathrm{Hz}$, $F\lambda/2h=89 \, \mathrm{Hz}$.

\item[{\bf Fig. \ref{fig:fig_movie}.}]
$V_{{\overline{X}}}/E_R=4.3(2)$, 
  $\theta = 1.013(1)\pi$, corresponding to $t_{1,2,3}=[602,789,166]\,\mathrm{Hz}$, $\gamma_{x,y}/h=[0.15,0.59]\,\mathrm{Hz}$, $\Delta_{AB}=266\,\mathrm{Hz}$. $F\lambda/2h=89 \, \mathrm{Hz}$.
  
\end{description}
 
\section{Lattice potential}\label{s:harmonictrap}

The phase $\varphi$ (see Eq. \ref{eqlattice})
between the $X$ and $Y$ beams at the position of the atoms is stabilized
interferometrically using a pair of additional
beams detuned from each other and from $\overline{X}, X$ and $Y$. This
results in a weak additional lattice along each axis of about $0.1E_R$.

The laser beams used to create the lattice potential
cause the atoms to experience a weak harmonic trap in all three spatial directions.
For a lattice with $V_{{\overline{X}},X,Y}/E_R=[4.0(2),0.28(1),1.8(1)]$, the trapping
frequencies are given by $\omega_{x,y,z}/2\pi=[17.6(1),$ $31.8(5),32.7(5)]\,\mathrm{Hz}$, as calibrated
from dipole oscillations of the cloud. In general, the trapping frequencies approximately scale
as $\omega_x\propto\sqrt{V_{Y}}$, $\omega_y\propto\sqrt{V_{\overline{X}}+(V_{X}V_{Y}/V_{\overline{X}})}$ and
$\omega_z\propto\sqrt{V_{\overline{X}}+(V_{X}V_{Y}/V_{\overline{X}})+1.24V_{Y}}$ with respect
to the intensities of the lattice beams.
The harmonic confinement causes the time it takes for the atomic cloud to reach the center of the B.Z. after a Bloch oscillation to be slightly larger than $T_B$.

\section{Momentum-space integration of transfer fractions}
\label{sec:integration}

In the following we describe how the total transfer $\xi$ for the entire atomic cloud in 3D is calculated from the momentum-dependent transfer $\xi_D(q_x)$ following the procedure presented in \cite{LimPRL12}. We assume a semi-classical expression for the energy of the particles
\begin{equation} 
 \epsilon(\mathbf{q},\mathbf{r})=\frac{\hbar^2 q_x^2}{2m_x}+\frac{\hbar^2 q_y^2}{2m_y}+\frac{\hbar^2 q_z^2}{2m_z}+\frac{1}{2}\left(m_x\omega_x^2 x^2+m_y\omega_y^2 y^2+m_z\omega_z^2 z^2\right),
\end{equation}
where the effective masses $m_{x,y,z}$ are obtained by expanding the dispersion relation around quasi-momentum $\mathbf{q}=0$ in the tight-binding regime and $\omega_{x,y,z}$ are the trapping frequencies in the three different directions. At zero temperature the expression for the integrated transfer fraction is then given by
\begin{equation}
\xi =
\frac{\int_{\epsilon(\mathbf{q},\mathbf{r})<\mu}\xi_D(q_x)d\mathbf{q}d\mathbf{r}}{\int_{\epsilon(\mathbf{q},\mathbf{r})<\mu} d\mathbf{q}d\mathbf{r}} =
\frac{96}{15\pi q_F}\int_0^{q_F}\xi_D(q_x)\left(1-\frac{q_x^2}{q_F^2}\right)^{5/2} dq_x.
\end{equation}
Here we used the fact that the transfer only depends on $q_x$, and denote the Fermi wave vector $q_F=\sqrt{2m_x\mu}/\hbar$ and the chemical potential $\mu$. For the semi-classical expression of the energy of the particles, the total atom number $N$ is related via $\mu=\hbar\overline{\omega}(6N)^{1/3}$, where $\overline{\omega}$ is the geometric mean of the trapping frequencies.



\begin{thebibliography}{}

\bibitem{CastroNetoRMP09} A.~H. Castro Neto, F. Guinea, N.~M.~R. Peres, K.~S. Novoselov, A.~K. Geim, Rev. Mod. Phys. {\bf 81}, 109 (2009)

\bibitem{HasanRMP10} M.~Z. Hasan, C.~L. Kane, Rev. Mod. Phys. {\bf 82}, 3045 (2010)

\bibitem{GomesNature12} K. K. Gomes, W. Mar, W. Ko, F. Guinea, H. C. Manoharan, Nature {\bf 483}, 306 (2012)

\bibitem{KuhlPRB10} U. Kuhl, S. Barkhofen, T. Tudorovskiy, H.--J. Stöckmann, T. Hossain, L. de Forges de Parny, F. Mortessagne, Phys. Rev. B {\bf 82}, 094308 (2010)

\bibitem{KlingPRL10} S. Kling, T. Salger, C. Grossert, M. Weitz, Phys. Rev. Lett. {\bf 105}, 215301 (2010)

\bibitem{SalgerPRL11} T. Salger, C. Grossert, S. Kling, M. Weitz,
Phys. Rev. Lett. {\bf 107}, 240401 (2011)

\bibitem{TarruellNature12} L. Tarruell, D. Greif, T. Uehlinger, G. Jotzu, T. Esslinger, Nature {\bf 483}, 302 (2012)

\bibitem{DahanPRL96} M. Ben Dahan, E. Peik, J. Reichel, Y. Castin, C. Salomon, Phys. Rev. Lett. {\bf 76}, 4508 (1996)

\bibitem{BreidNJP06} B. M. Breid, D. Witthaut, H. J. Korsch, New J. Phys. {\bf 8}, 110 (2006)

\bibitem{StrableyPRA06} J. Sebby-Strabley, M. Anderlini, P. S. Jessen, J. V. Porto, Phys. Rev. A {\bf 73}, 033605 (2006)

\bibitem{FoellingNature07} S. Fölling, S. Trotzky, P. Cheinet, M. Feld, R. Saers, A. Widera, T. Müller, I. Bloch, Nature {\bf 448}, 1029 (2007)

\bibitem{SalgerPRL07} T. Salger, C. Geckeler, S. Kling, M. Weitz, Phys. Rev. Lett. {\bf 99}, 190405 (2007)

\bibitem{WirthNaturePhys11} G. Wirth, M. Ölschläger, A. Hemmerich, Nature Phys. {\bf 7}, 147 (2011)

\bibitem{SoltanPanahiNaturePhys11} P. Soltan-Panahi, J. Struck, P. Hauke, A. Bick, W. Plenkers, G. Meineke, C. Becker, P. Windpassinger, M. Lewenstein, K. Sengstock, Nature Phys. {\bf 7}, 434 (2011)

\bibitem{JoPRL12} G.-B. Jo, J. Guzman, C. K. Thomas, P. Hosur, A. Vishwanath, D. M. Stamper-Kurn, Phys. Rev. Lett. {\bf 108}, 045305 (2012)

\bibitem{KoehlPRL05} M. K\"ohl, H. Moritz, T. St\"oferle, K. G\"unter, T. Esslinger, Phys. Rev. Lett. {\bf 94}, 080403 (2005)

\bibitem{KastbergPRL95} A. Kastberg, W. D. Phillips, S. L. Rolston, R. J. C. Spreeuw, P. S. Jessen, Phys. Rev. Lett. {\bf 74}, 1542 (1995)

\bibitem{Esslinger10} T. Esslinger, Annu. Rev. Condens. Matter Phys. {\bf 1}, 129 (2010)

\bibitem{MontambauxEPJ09} G. Montambaux, F. Piéchon, J.-N. Fuchs, M.O. Goerbig, Eur. Phys. J. B {\bf 72}, 509 (2009)

\bibitem{LimPRL12} L.-K. Lim, J.-N. Fuchs, G. Montambaux, Phys. Rev. Lett. {\bf 108}, 175303 (2012)

\bibitem{StueckelbergHPA32} E. C. G. Stückelberg, Helv. Phys. Acta {\bf 5}, 369 (1932)

\bibitem{ShevchenkoPR10} S. N. Shevchenko, S. Ashhab, F. Nori, Phys. Rep. {\bf 492}, 1 (2010)

\bibitem{HasegawaArXiv12} Y. Hasegawa, K. Kishigi, arXiv:1207.6841v1

\bibitem{Park} T. J. Park, J. C. Light, J. Chem. Phys. {\bf 85}, 587 (1986)

\end{thebibliography}
\end{document}